\documentclass[twoside,slac_one]{revtex4}
\usepackage{graphicx}
\usepackage{fancyhdr}
\usepackage{amsmath} 
\usepackage{bm}
\usepackage{amsxtra}
\usepackage{amssymb}
\usepackage{amsthm}
\usepackage{latexsym}
\usepackage{lscape}
\usepackage{xspace}

\newcommand{\met}{\ensuremath{E_{\mathrm{T}}^{\mathrm{miss}}}\xspace}
\newcommand{\pt}{\ensuremath{p_{\mathrm{T}}}\xspace}

\newcommand{\wboson}{\ensuremath{W}\xspace}
\newcommand{\zboson}{\ensuremath{Z}\xspace}
\newcommand{\rjets}{\ensuremath{R_{\mathrm{jet}}}\xspace}
\newcommand{\pythia}{{\sc Pythia}\xspace}
\newcommand{\alpgen}{{\sc Alpgen}\xspace}

\newcommand{\powheg}{{\sc Powheg}\xspace}

\def\GeV{\ifmmode {\mathrm{\ Ge\kern -0.1em V}}\else
                  \textrm{Ge\kern -0.1em V}\fi}%
\def\MeV{\ifmmode {\mathrm{\ Me\kern -0.1em V}}\else
                  \textrm{Me\kern -0.1em V}\fi}%
\def\TeV{\ifmmode {\mathrm{\ Te\kern -0.1em V}}\else
                  \textrm{Te\kern -0.1em V}\fi}%

\setcitestyle{numbers}
\begin{document}

\title{A Measurement of the Ratio of the W + 1 Jet to Z + 1 Jet Cross Sections with ATLAS}

%

\author{A. Meade and B. Brau on behalf of the ATLAS Collaboration}
\affiliation{Department of Physics, University of Massachusetts, Amherst, MA, USA}

\begin{abstract}
The measurement of hadronic activity recoiling against \wboson and \zboson vector
bosons provides an important test of perturbative QCD, as well as a
method of searching for new physics in a model independent fashion. We
present a study of the cross-section ratio for the production of \wboson and
\zboson gauge bosons in association with exactly one jet 
$\rjets =
\sigma(\wboson+\mathrm{1jet})/\sigma(\zboson+\mathrm{1jet})$,
in $pp$ collisions at $\sqrt{s}$ = 7 \TeV. The study is
performed in the electron and muon channels with data collected with
the ATLAS detector at the LHC. The ratio \rjets is studied as a
function of the cumulative transverse momentum distribution of the
jet. This result can be compared to NLO pQCD calculations
and the prediction from LO matrix element + parton shower
generators.
\end{abstract}

\maketitle

\thispagestyle{fancy}


\section{Introduction}
Measurements of the cross sections of the \wboson and \zboson bosons in association with hadronic
activity are important tests of the standard model, particularly
perturbative quantum chromodynamics. In addition many
searches for models of physics beyond the Standard Model have
significant $W$+jets or $Z$+jets backgrounds, adding to the importance of
understanding these processes. Individual measurements of \wboson or \zboson cross
sections in association with jets are limited by systematics that are
shared between the two measurements, such as the luminosity and the
jet energy scale. In contrast, in measurements of the ratio between
\wboson and \zboson cross sections, many of these uncertainties cancel due to the similar
nature of \wboson and \zboson production. For this reason the $W/Z$ cross
section ratio can be the basis for a program of high precision measurements.

This study reports on a measurement of the ratio of production
cross sections of the \wboson and \zboson bosons with exactly one associated jet
with $\pt > 30 \GeV$ as a function of jet transverse momentum
threshold. This binning implies that each bin contains all events with a
jet above the threshold. The data for this analysis consists of the entire 2010 ATLAS dataset,
operating at the Large Hadron Collider (LHC). This corresponds to 33
pb$^{-1}$ of integrated luminosity, taken at a center of mass energy
of 7 \TeV. A paper on this measurement has been produced by the ATLAS
collaboration \cite{RjetsPreprint}.

The ratio is measured in a fiducial measurement volume
of the detector where the leptons and jets are measured with good
resolution, in order to avoid theoretical uncertainties associated
with modeling particles outside of the detector volume. The fiducial
region is defined by the following kinematic ranges of the lepton(s) $\ell$, neutrino
$\nu$, and jet: $\pt^{\mathrm{\ell}} > 20~
\GeV$, $\pt^{\rm \nu}>25 \GeV$, $\pt^\mathrm{jet} > 30 \GeV$,
$|\eta^\mathrm{jet}| < 2.8$. The electrons and muons were restricted
to different pseudorapidity ranges, for electrons $1.52 <
|\eta| < 2.47$ or $|\eta| < 1.37$, while for muons $|\eta|<2.4$. For the W, the
transverse mass, defined as $m_{\rm T} =\sqrt{2\ensuremath{p_{\mathrm{T}}}^{\ell}\ensuremath{p_{\mathrm{T}}}^{\nu}(1-\cos(\phi^{\ell}-\phi^{\nu}))}$
was required to satisfy $m_{\rm T} > 40 \GeV$.  The dilepton invariant mass
of the \zboson was required to be within the range $71 < m_{\rm \ell\ell} <
111\GeV$.

Particle level jets were defined as jets reconstructed in simulated events
by applying the anti-$k_{\rm{t}}$ jet reconstruction
algorithm \cite{Cacciari:2008gp} with a radius parameter $R = 0.4$ to all
final state particles with a lifetime longer than 10~ps (including muons and
non-interacting particles).  Particle level electrons were defined by
including the energy of all radiated photons within a cone of $\Delta R =
0.1$ around each electron (with $\Delta R = \sqrt{(\Delta\eta)^2 +
  (\Delta\phi)^2}$). In the muon case the ``bare'' final state muon is
taken as the particle level muon, with no radiated photons included.

\section{The ATLAS Detector}
The ATLAS detector is described in detail
elsewhere \cite{AtlasDetector}. Immediately surrounding
the interaction point is the inner detector (ID),
providing precision tracking and vertexing capabilities, surrounded by
a large solenoidal magnet. Outside of this volume lies the
electromagnetic and hadronic calorimeters, used for energy measurement
and electron and photon identification. The outermost portion of the
detector is the muon spectrometer (MS), based on three large
superconducting toroids and a system of three stations of trigger
chambers and precision tracking chambers.

\section{Simulated Event Samples}
\label{sec:simulation}

Simulated event samples were used to correct signal yields for
detector effects, for some of the background estimates, and for comparison
of the results to theoretical expectations. Samples of $W \rightarrow
\ell\nu+N_\mathrm{parton}$ and $Z \rightarrow \ell\ell+N_\mathrm{parton}$ (where
$\ell=e,\mu,\tau$) were generated using
\alpgen~\cite{Alpgen}. Background and additional signal samples were
generated with \pythia~\cite{Pythia} and \powheg~\cite{Powheg}. To take
into account overlapping pp collisions some samples were generated
with multiple non-diffractive scattering events overlapping with the
primary collision event. These samples were reweighted such that the
distribution of the number of reconstructed primary vertices matched that of the
data sample. GEANT4~\cite{Geant4} was used to simulate the detector
response for all samples, and these samples were subject to the same reconstruction
and analysis chain as the $pp$ collision data.

Predictions for the \wboson and \zboson cross sections as a function of jet \pt
threshold at NLO were obtained using MCFM, with corrections to
particle level calculated using \pythia, to account for initial and
final state radiation, hadronization, and underlying event.

\section{Data and Event Selection}
\label{sec:selection}

Kinematic requirements for reconstructed muon and electron
candidates match those of the fiducial definition: $p_{T} >
25\GeV$, $\left|\eta\right| < 2.4$. In the electron channel events
were triggered based on the presence of an electromagnetic cluster with $E_{T} > 15\GeV$. Electron candidates
were required to satisfy lateral shower containment, shape and width
criteria, and minimal leakage into the hadronic calorimeter in order
to be classified as ``medium''. Consistency between track p$_{T}$
and cluster energy as well as tighter hit requirements are
additionally required to categorize an electron as
``tight''.  For the electron channel the pseudorapidity range region
$1.37 < \left|\eta\right| < 1.51$ is rejected due to a gap in the calorimeter coverage.

In the muon channel, events were selected by the trigger based on muon
spectrometer hits consistent with a track of 10 or 13 \GeV, the
tighter requirement applying to later data with higher instantaneous luminosity. Muon candidates were required to have
inner detector and muon spectrometer tracks and pass kinematic
requirements to be classified as ``medium''. To be classified as
``tight,'' consistency is demanded between ID and MS track, and requirements on the
number of inner detector hits were made. In addition, the impact
parameters of the muon are required to be consistent with the
interaction point, and the inner detector track is required to be
isolated from other energetic tracks, with $\sum{p_T}$ of all tracks
within $\Delta R < 0.2$ less than 1.8 \GeV.

Each event was required to have a reconstructed primary vertex with
three or more tracks, and to be within 150 mm of the center of the
detector along the beamline. \wboson candidates were required to have exactly one ``tight''
lepton with no additional ``medium'' leptons, $\met > 25 \GeV$,
and $M_T > 40 \GeV$. \zboson candidates are required to have exactly
two leptons of opposite charge, one ``tight'', and one of ``medium'' or higher quality. The invariant
mass $m_{\ell\ell}$ of the lepton pair is required to be in the range
$71 < m_{\rm \ell\ell}<
111$ \GeV.

Reconstructed jets were defined using the anti-$k_{\rm{t}}$ jet reconstruction
algorithm with a radius parameter $R = 0.4$ as
in the fiducial definition. These jets were required to be within the kinematic range 
$p_{T} >30 \GeV$, $\left|\eta\right| < 2.8$. Events were rejected if identified as likely to
contain a jet from detector noise or a cosmic ray. Events with jets within a
cone of $\Delta R < 0.6$ from an electron are rejected to avoid
distortion of the shower shape from the nearby electromagnetic
shower, which could result in reconstruction inefficiency or incorrect energy measurement. Jets from additional
interactions in a bunch crossing were rejected by requiring that 75\%
of the scalar sum $p_{T}$ of all tracks associated with a jet came from
tracks originating from the same primary vertex. This approach is termed
the jet vertex fraction (JVF) algorithm. Only events containing
exactly one jet passing these requirements were selected as candidate
events.

Using these $W$($Z$) selection criteria, 12112 (948) events and
12995 (1376) events were found in the electron and muon channels
respectively.

\section{Background Estimation}
Backgrounds to the measurement are determined independently for each
bin of the measurement in jet $p_T$ threshold. These backgrounds are
categorized as resulting from electroweak or QCD multijet processes, and are
subtracted from the total number of events passing selection using the following
formula to obtain the signal event yield $N_\mathrm{sig}$:

\begin{eqnarray}
N_\mathrm{sig} & = &   N_\mathrm{tot} \cdot (1-f_\mathrm{multijet})   (1-f_\mathrm{ewk})
\end{eqnarray}

The predicted number of events passing selection
cuts is shown for all signal and background processes in Table~\ref{tbl:backgrounds}.

The electroweak background fraction $f_{\mathrm{ewk}}$ is estimated from
simulated data as a fraction of the total multijet subtracted
event count. This has the advantage of having no dependence on the
luminosity, and reduces dependence of the background on our
measurement volume, as the electroweak processes have similar
kinematic distributions.

The multijet background processes are estimated using data driven
methods that vary based on the channel. In the $W \rightarrow e \nu$
channel, signal and background templates were fit to the control
region $15 \GeV < \met < 55 \GeV$ to determine the background fraction. The
signal and electroweak background template shapes are taken from
simulation. The multijet event shapes are found from data by selecting ``medium''
electron candidates and requiring that 2 of the ``tight'' selection
requirements, as well as isolation, are failed, producing what is essentially
an electron-like multijet sample. The $Z \rightarrow ee$ background
estimate was performed using a similar method by fitting templates to the
dilepton invariant mass spectrum. Because of smaller background, a
looser electron selection and reversal of two of the ``medium''
selection criteria were used in the determination of the multijet
template.

In the  $W \rightarrow \mu \nu$ channel, the multijet background is
estimated by measuring the number of events passing all selection, and
all selection but isolation. The efficiency of this selection is then
measured on data control samples in the data. Dimuon events from
\zboson events are used to measure the efficiency for muons from
electroweak processes like the \wboson and \zboson signal, and
dijet events are used to measure the efficiency for muons from the
multijet background. Since these two efficiencies are significantly different,
the multijet background fraction can be determined. In the $Z
\rightarrow \mu \mu$ channel, the multijet background is very
small. In this case a scale factor is derived by comparing
non-isolated dimuon pairs in data and simulation, and used to
normalize a simulated multijet sample.

In the electron \wboson channel, multijet backgrounds are the dominant
background due to the relatively high probability of a jet fragmentation product being
reconstructed as an electron. The electron
multijet background fraction was about 16\% for the lowest jet threshold, 30
\GeV. $Z \rightarrow ee$, $W \rightarrow \tau \nu$, and $t
\bar{t}$ events are all significant sources of electroweak background,
which is around 3.4\%. The dominant backgrounds in the $W \rightarrow
\mu \nu$ channel are $W \rightarrow \tau \nu$ decaying to a muon (2\%), $Z
\rightarrow \mu \mu$ with one muon failing to be reconstructed (3\%), and
multijet events (3\%). In both the electron and muon \zboson channels,
backgrounds were very small, less than $1\%$ for all \zboson candidates.

\begin{table*}[!htbp]
  \begin{center}
    \begin{tabular}{|l|c|c||l|c|c|}
      \hline
      Process & $W\rightarrow e \nu$  & $Z\rightarrow e e$  &
      Process & $W\rightarrow \mu \nu$  & $Z\rightarrow \mu\mu$  \\
      \hline 
      $W\rightarrow e \nu$   & $9340 \pm 40 $ & $3 \pm 1$ &
      $W \rightarrow \mu \nu$ & $11860 \pm 40$ & $ 4 \pm 2$ \\
      $Z \rightarrow ee$   & $106 \pm 3$ & $880 \pm 10$ &
      $Z \rightarrow \mu \mu$ & $360 \pm 6 $ & $1370 \pm 40$ \\
      $W\rightarrow \tau \nu$  & $191 \pm 6$ & $ 0.2 \pm 0.2 $ &
      $W \rightarrow \tau \nu$ & $234 \pm 6$ & $0.3 \pm 0.6$ \\
      $t\bar{t}$ & $33\pm 1$ & $1.9 \pm 0.2$ &
      $t \bar{t}$ & $35 \pm 1$ & $3 \pm 2$ \\
      $Z \rightarrow \tau\tau$  & $ 19 \pm 1 $ & $0.3 \pm 0.1$&
      $Z \rightarrow \tau \tau$ & $22 \pm 1$ & $0.3 \pm 0.6$\\
      Multijet  & $1800 \pm 60$ & $2.9 \pm 0.6$ &
      Multijet  & $380 \pm 70$ & $4 \pm 4$ \\
      Other & $58 \pm 2$  & $1.1 \pm 0.1$ &
      Other & $117 \pm 1$ & $8 \pm 3$ \\
      \hline
      Total & $11550 \pm 70$ & $ 880 \pm 10$ &
      Total & $13010 \pm 80$ & $1380 \pm 40$ \\
      \hline
      Data $N_\mathrm{tot}$      &  $12112$  &    $948$   &
      Data  $N_\mathrm{tot}$ & $12995 $ & $1376$ \\
      \hline
    \end{tabular}
  \end{center}

  \caption{Predicted  and observed events in data in the
electron and muon channels for the $W$ and $Z$ selections for $\int
\mathcal{L} dt =
33$~$pb^{-1}$.  Background estimates are quoted for a jet $p_T$ threshold of
$30 \GeV$.  Only statistically errors are displayed. ``Other'' includes contributions from diboson and single top events.}
\label{tbl:backgrounds}
\end{table*}

\section{Correction Procedure}
\label{sec:signal}
Event yields were corrected for trigger efficiency
($\epsilon_{\mathrm{trig}}^{\mathrm{\ell}}$), lepton identification efficiency
($\epsilon^{\mathrm{\ell}}$), and differences between boson acceptance at
detector level ($A_{\mathrm{reco}}$) and particle level ($A_{\mathrm{part}}$) due to resolution effects
($C_{\mathrm{V}}^{\mathrm{\ell}} = \frac{A_{\mathrm{reco}}}{A_{\mathrm{part}}}$). The number of signal events for each boson at
particle level was then obtained using
\begin{eqnarray}
\label{eqn:e-had}
{N_\mathrm{part}^{\rm \ell,V}} &=& \frac{N_\mathrm{sig}^{\rm \ell,V}} { \epsilon_{\mathrm{trig}}^{\mathrm{\ell}} \times \epsilon^{\mathrm{\ell}}
\times C_{\mathrm{V}}^{\mathrm{\ell}}}.
\end{eqnarray}
where the boson corrections $C_{\mathrm{V}}^{\mathrm{\ell}}$ correct the observed phase space to
the truth (particle) level kinematic phase space, accounting for the resolution of
leptons and \met.

Trigger and identification efficiency for leptons were determined from
an unbiased control sample by selecting a well identified ``tag''
lepton and a additional ``probe'' inner detector track in $Z
\rightarrow \ell \ell$ candidate events. The efficiency is the
fraction of probe tracks matched to a reconstructed lepton
and passing all identification requirements. These efficiencies were found to
be largely independent of jet kinematics, other than
through correlations with lepton kinematics due to the jet
recoiling against the \wboson or \zboson. Therefore, these
efficiencies were binned in lepton \pt and $\eta$ for the electron
channel, and $\eta$ and $\phi$ for the muon channel, for all jet
\pt thresholds.

When the ratio of cross sections is measured, jet resolution and scale
effects almost completely cancel. A correction $C_{jet}^{\rm \ell}$
was applied to take into account small differences. This correction is
defined as the
ratio between the \rjets value as a function of particle level jet \pt threshold
and the \rjets value as a function of reconstructed jet \pt threshold. This
factor is measured in simulation samples and applied as follows:

\begin{eqnarray}
\rjets &=&
\frac{
N_\mathrm{part}^{\rm \ell,W}
}{
N_\mathrm{part}^{\rm \ell,Z}
}
\times C^{\rm \ell}_\mathrm{jet}.
\end{eqnarray}

\begin{figure}
\centering
\includegraphics[width=0.45\textwidth]{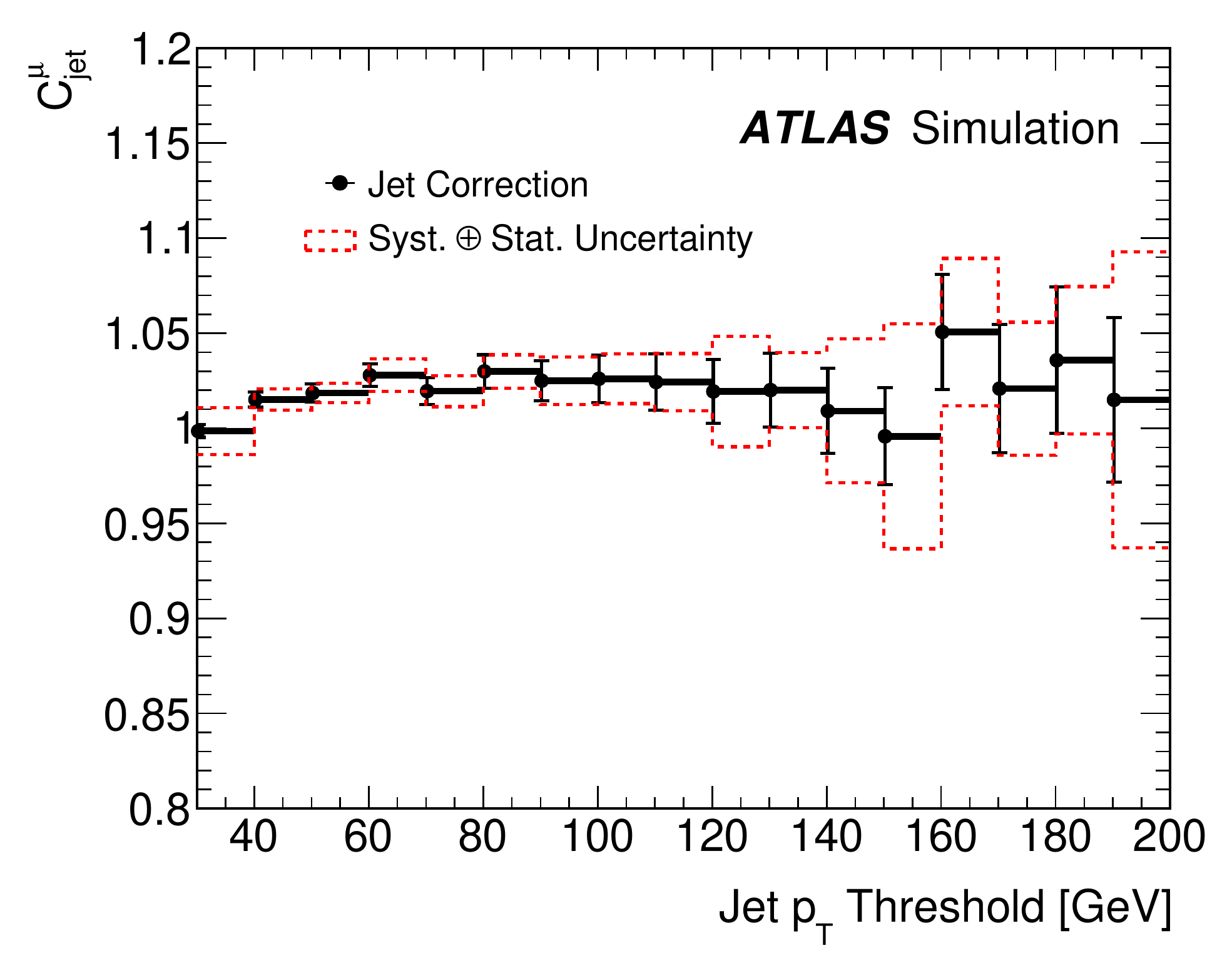}
\caption{
Detector jet spectrum correction ($C_{\mathrm{jet}}^{\rm \mu}$) on \rjets in the
muon channel derived from \alpgen. The uncertainty, shown as a dashed red
line accounts for the difference between \pythia and \alpgen
generators. The black error bars show the statistical uncertainty alone.}
\label{fig:jet-correction}
\end{figure}

This correction is shown for the muon channel in Figure
\ref{fig:jet-correction}. Differences from unity correspond to an offset in the amount of
migration between jet \pt bins between W+jet and Z+jet events. This is
mostly due to selection differences in the two channels before the jet
selection, such as the requirement that the decay products be located
in the detector volume. Note that binning in terms of jet \pt
threshold eliminates migration across the upper bin edges.

\section{Systematic Uncertainties}

\begin{figure*}[!htbp]
  \begin{minipage}{0.49\linewidth}
  \begin{center}
    \includegraphics[width=0.95\linewidth]{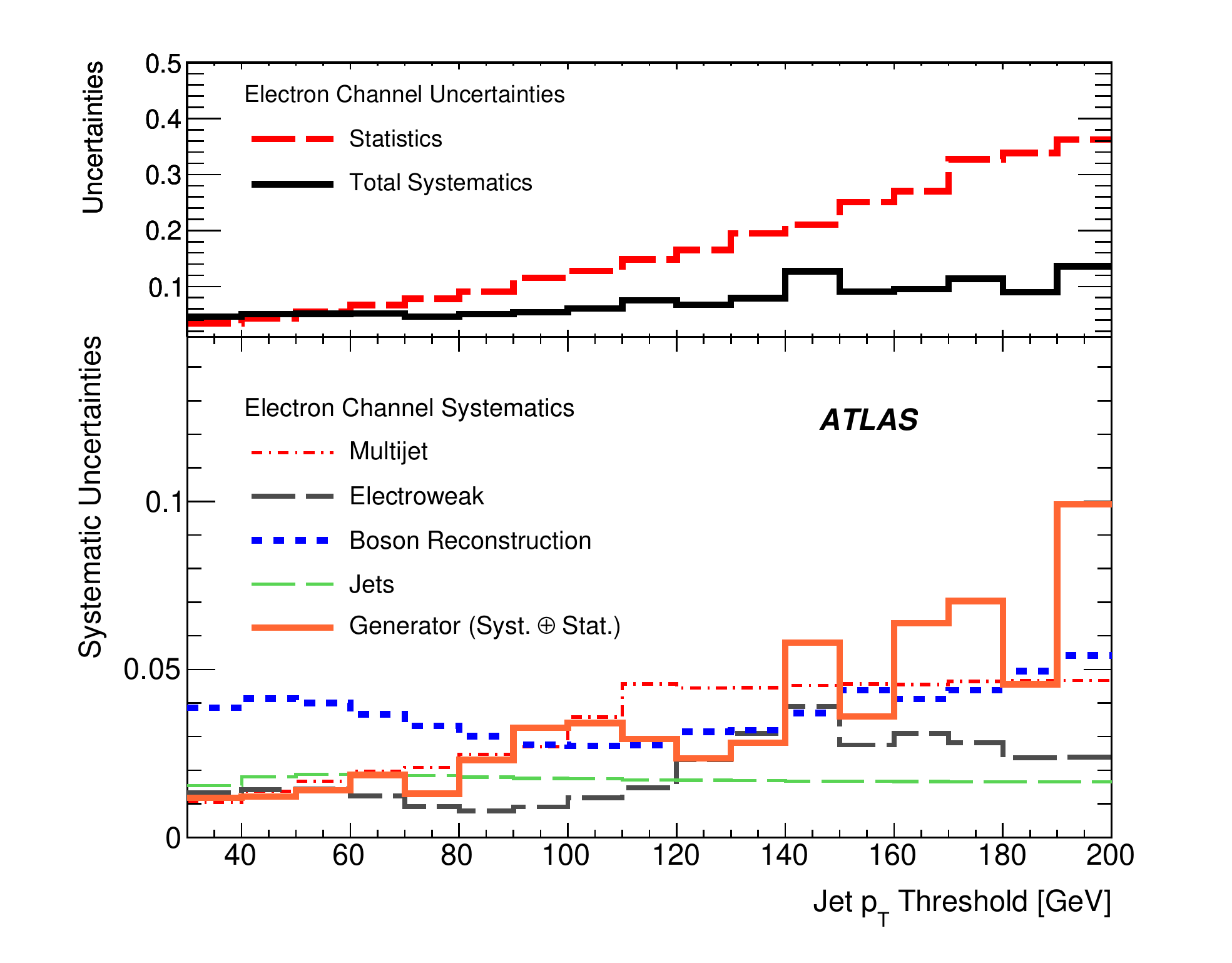}
\end{center}
\end{minipage}
  \begin{minipage}{0.49\linewidth}
    \begin{center}
    \includegraphics[width=0.95\linewidth]{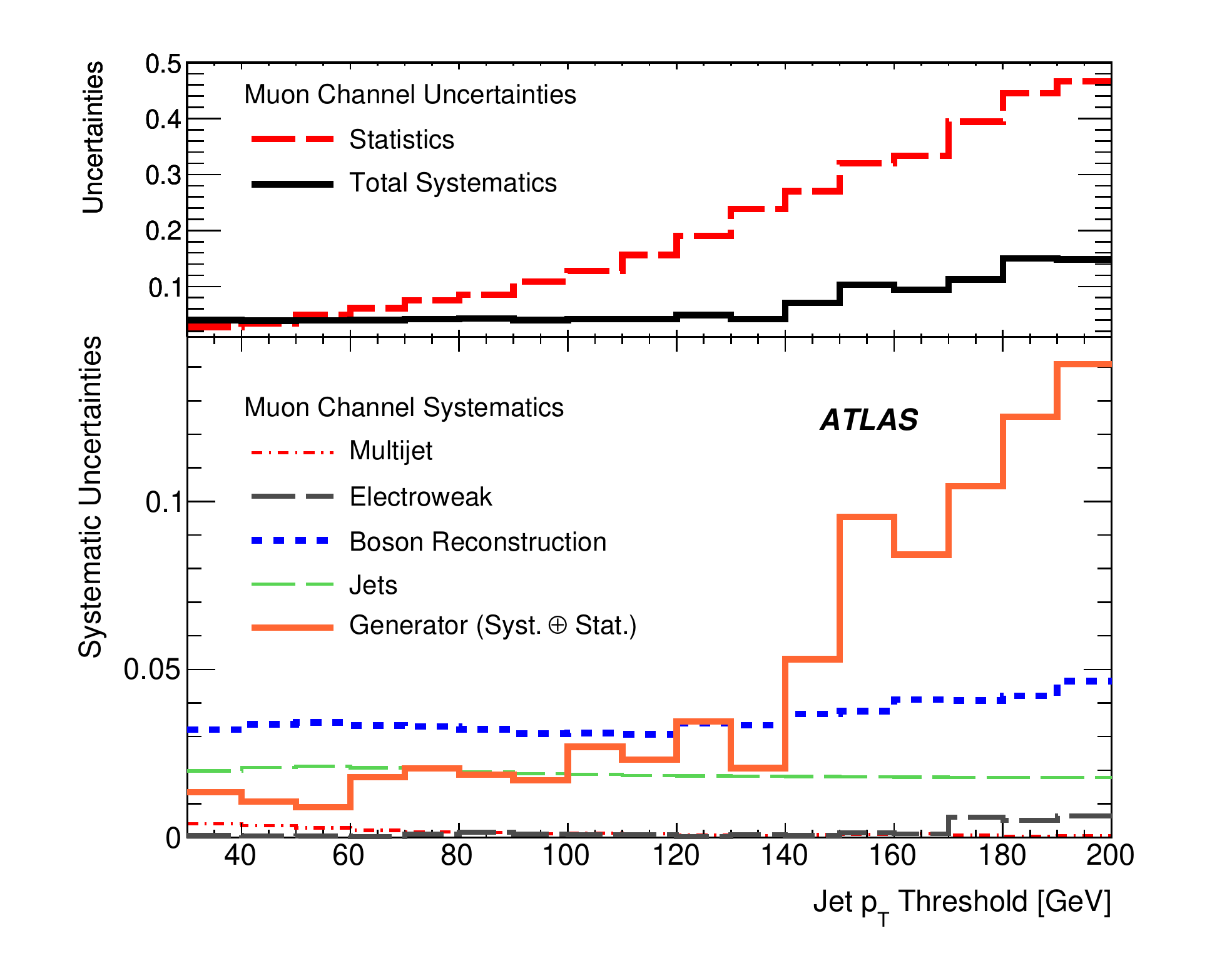}
  \end{center}
  \end{minipage}
\caption{
Relative systematic uncertainties on \rjets in the electron channel (left)
and in the muon channel (right). The top plot displays the total systematic
and statistical uncertainty (shown as red dashed line) versus jet \pt
threshold. The lower plot shows the breakdown of the systematic
uncertainties. The ``boson reconstruction'' entry contains the uncertainties related to
the leptons and \met (including trigger and lepton
identification). The ``jet'' entry contains systematics of the jet
correction as well as the jet energy scale and
resolution. Uncertainties from each group were added in quadrature.}
\label{fig:systematics}
\end{figure*}

In order to properly take into account correlations between the \wboson and
Z measurements, and to maximize cancellation of systematic
uncertainties, all such uncertainties were measured as a relative
change in \rjets. The total systematic uncertainty varies from 4\%
to 15\%, generally increasing with increased jet \pt
threshold. Systematic uncertainties on the measured ratio were divided into
five broad groups: multijet background, electoweak background, boson reconstruction
(combining $\epsilon_\mathrm{trig}^{\rm \ell} \times \epsilon^{\rm \ell} \times
C_{\rm V}^{\rm \ell}$), the jet spectrum correction $C_{\rm jet}^{\rm \ell}$, and uncertainties
related to generator differences, as shown in Figure \ref{fig:systematics}. 

Systematics on the multijet background were estimated by varying the
parmeters of each method, such as the selection of control samples
used to derive the background fractions. Systematics on the electoweak
background are estimated by varying the lepton identification
criteria, by varying momentum scale and resolution, and by comparing
samples generated with and without additional overlapping pp interactions. 

Boson reconstruction systematic uncertainties included those
associated with identification and trigger efficiency, as well as
scale and resolution effects. The single lepton trigger and identification efficiency uncertainties were
derived using the methods documented in Ref.~\cite{Aad:2010yt}, and
then propagated to uncertainties on \rjets. This leads to a total
identification efficiency uncertainty of 1.1\%(1.7\%) for electrons (muons)
independent of jet \pt. Trigger efficiency uncertainties were very small in both
channels.

Uncertainties involving the scale of lepton momenta and
\met were evaluated by directly scaling up and down the quantity in question by
the measured scale uncertainty. In each case the deviation of \rjets was applied
as a systematic uncertainty. Similarly, resolution effects were
evaluated by smearing the quantities by a gaussian with the width of
the measured resolution, repeating this process and taking the root mean squared deviation as
the systematic uncertainty. Lepton momentum scale and
resolution measurements and their uncertainties are determined by
comparing the width and position of the invariant mass spectrum of Z candidates in
data to that in a simulated sample.

Jet related systematics included those involved with jet energy
scale (JES), jet energy resolution (JER) and their associated
uncertainties. These quantities
were determined from data and simulation comparisons~\cite{JES}.  The JES
uncertainty includes components from calibration and jet sample composition
differences.  The JES calibration uncertainty varies with $|\eta|$ and \pt,
and ranges from 4\% to 8\%.  The JES and JER were measured with di-jet
events, which have different proportions of quark and gluon initiated jets
than events containing vector bosons.  Therefore, an uncertainty was assigned
to account for the difference in calorimeter response between jets in $V$ +
jet events and the di-jet events used for calibration, ranging from 2 to
5\%, and was added in quadrature to the JES calibration uncertainty.  The
total JES uncertainty ranges from approximately 10\% at 20 \GeV to 5\% at
100 \GeV. The JER is measured to vary from 2\% at 20 \GeV to 1\% at
100 \GeV. The uncertainties on \rjets due to JER and JES were evaluated using the same
approach as for the lepton scale and resolution uncertainties. The uncertainties on
\rjets due to the JER and JES were found to be approximately 0.5\% and 2\%
respective. This uncertainty includes a very small term to to take into account the
uncertainty on JER itself.

Jet related systematics also included the dominant uncertainty related to multiple pp collisions, the
uncertainty of the efficiency of the JVF algorithm. Applying this
algorithm to simulation with multiple interactions was found to produce
consistent results with simulation not including additional
interactions. The residual difference on \rjets between these cases
was used as a systematic uncertainty on  $C_{\rm jet}^{\rm \ell}$.

To account for systematics associated with generator modeling, correction
factors were computed with samples generated with \pythia instead of
\alpgen, and the observed difference was applied as a systematic uncertainty.  Systematic
uncertainties were assigned from this variation to the following
corrections: ($C_\mathrm{jet}^{\rm \ell}$), the boson reconstruction correction
$C_{\rm V}^{\rm \ell}$, and the electroweak background estimation $f_\mathrm{ewk}$.
At large jet \pt threshold, where the statistical uncertainty on the
measurement dominates the total uncertainty, this systematic uncertainty is
limited in statistical precision due to the size of the samples used, and is the dominant systematic
uncertainty.

Systematic uncertainties were also estimated on the theoretical
prediction. As our NLO parton-level calculation does not include the effects of
hadronisation and underlying event, a correction was computed using
\pythia as a function of jet \pt threshold. The uncertainty on this
correction was evaluated by observing the effect of various generator
tunes on the final \rjets result. These tunes increased or decreased
the amount of underlying event, or varied the parameters controlling initial and final state
radiation. Renormalization and factorization scale uncertainties were
also included in our systematic uncertainties, as were PDF
uncertainties.

\section{Results}

The ratio \rjets was measured in the fiducial region of the ATLAS detector
as a function of jet \pt threshold, and corrected for
detector effects. The electron and muon measurements were performed in
slightly different phase space, due to the different $\eta$ range and
electron-jet isolation requirements, as well as for the different QED
treatment between electron and muon definitions.  The observed signal
yields were corrected to recover the yield at particle level as
described in Section~\ref{sec:signal}.

\begin{figure*}[!htbp]
  \begin{minipage}{0.49\linewidth}
  \begin{center}
    \includegraphics[width=0.95\linewidth]{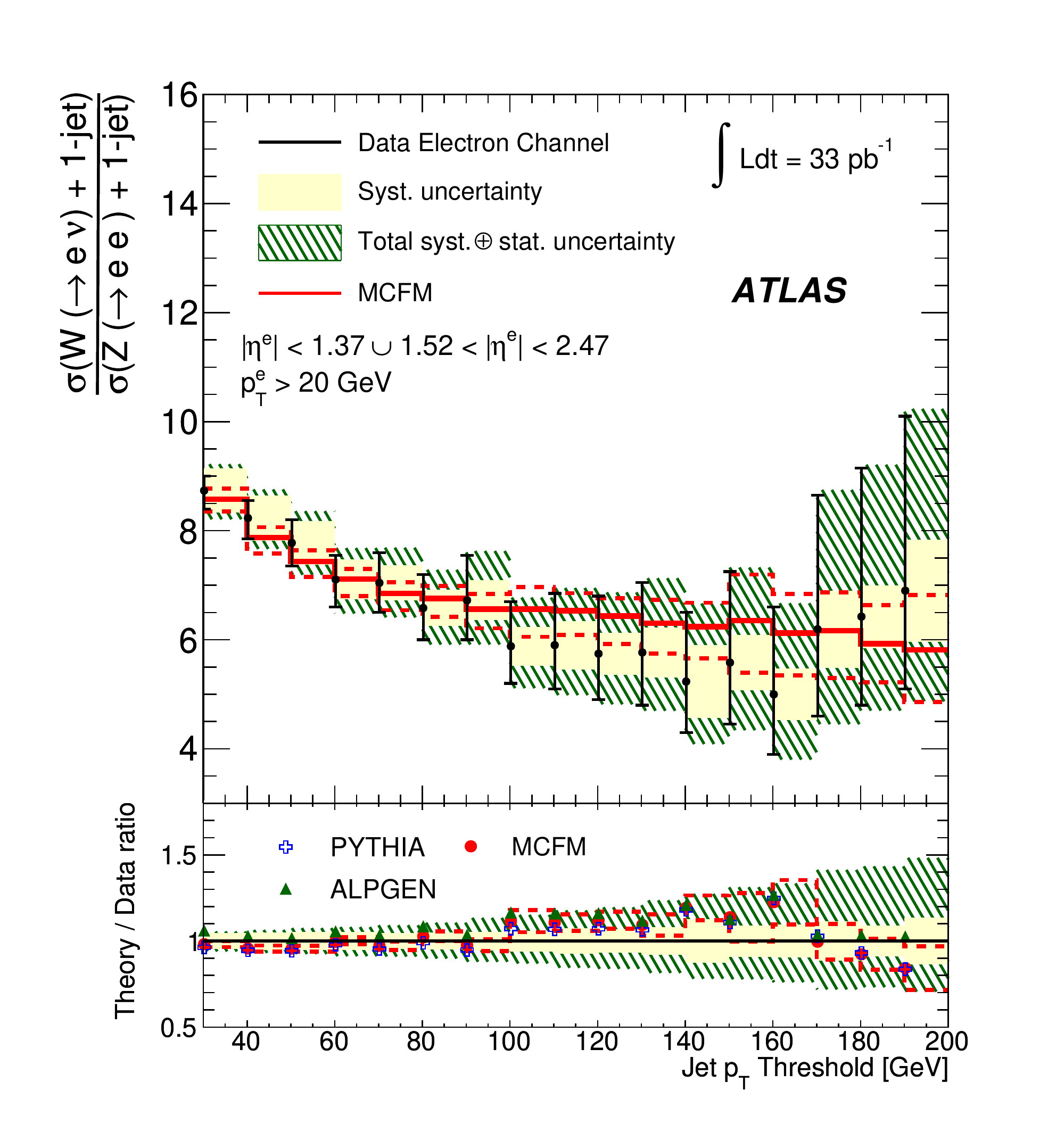}
  \end{center}
  \end{minipage}
  \begin{minipage}{0.49\linewidth}
    \begin{center}
    \includegraphics[width=0.95\linewidth]{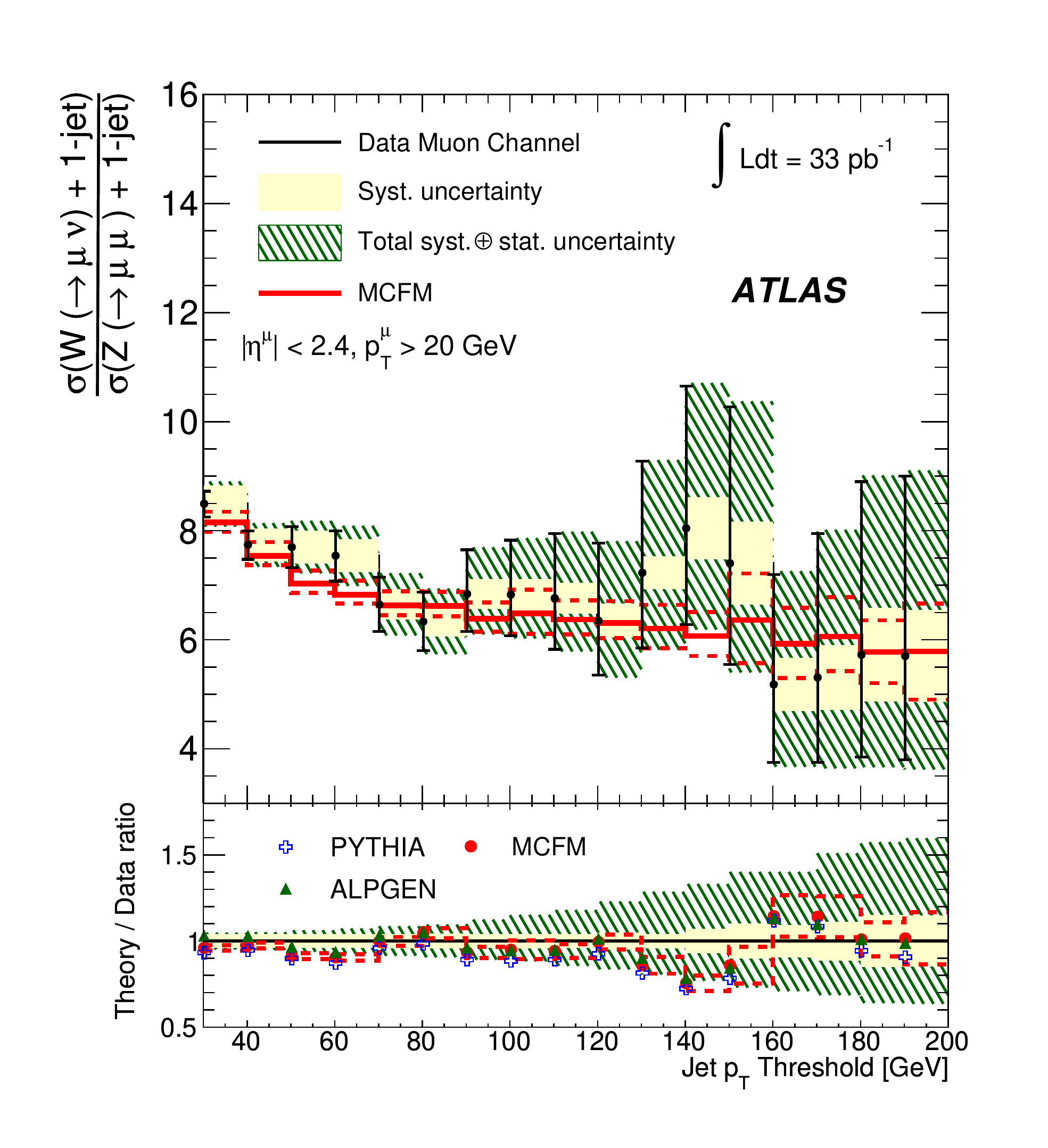}
  \end{center}
  \end{minipage}
\caption{
Results for \rjets in the electron channel (left) and in the muon channel
(right) for their respective fiducial regions.  The results are compared to
NLO predictions from MCFM (corrected to particle level using \pythia). Data
are shown as black points at the lower bin edge corresponding to the jet \pt
threshold with black error bars indicating the statistical uncertainties.
The yellow band shows all systematic uncertainties added in quadrature and
the green band shows statistical and systematic uncertainties added in
quadrature. The theory uncertainty (dashed line) shown on the MCFM
prediction includes uncertainties from PDF and renormalization and
factorization scales.  Note that these threshold data and their associated
uncertainties are correlated between bins.}
\label{fig:results}
\end{figure*}

The corrected ratio \rjets of the production cross sections in the leptonic
(electron or muon) decays of the gauge bosons $W$ and $Z$ in association
with exactly one jet is shown in Figure~\ref{fig:results} as a function of
the jet \pt threshold for the electron (left) and muon (right) channels.
As the jet \pt threshold increases, the ratio \rjets is expected to decrease
as the effective scale of the interaction becomes large compared to the
difference in boson masses.  This dependence is observed in the data.
The values for the lowest jet \pt threshold of $30 \GeV$ are:

\begin{eqnarray*}
\rjets(e)&=& 8.73 \pm 0.30 \,\mathrm{(stat)} \pm 0.40 \,\mathrm{(syst)}  \\
\rjets(\mu)&=& 8.49 \pm 0.23 \,\mathrm{(stat)} \pm 0.33 \,\mathrm{(syst)}
\end{eqnarray*}	
Statistical uncertainties were evaluated by repeating the measurement with
Monte Carlo pseudo-experiments assuming Poisson distributed data with a mean
at the observed yield.  Both electron and muon channel
results are individually compatible with the theoretical predictions.

\begin{figure*}[!htbp]
  \begin{minipage}{0.49\linewidth}
  \begin{center}
  \includegraphics[width=0.95\linewidth]{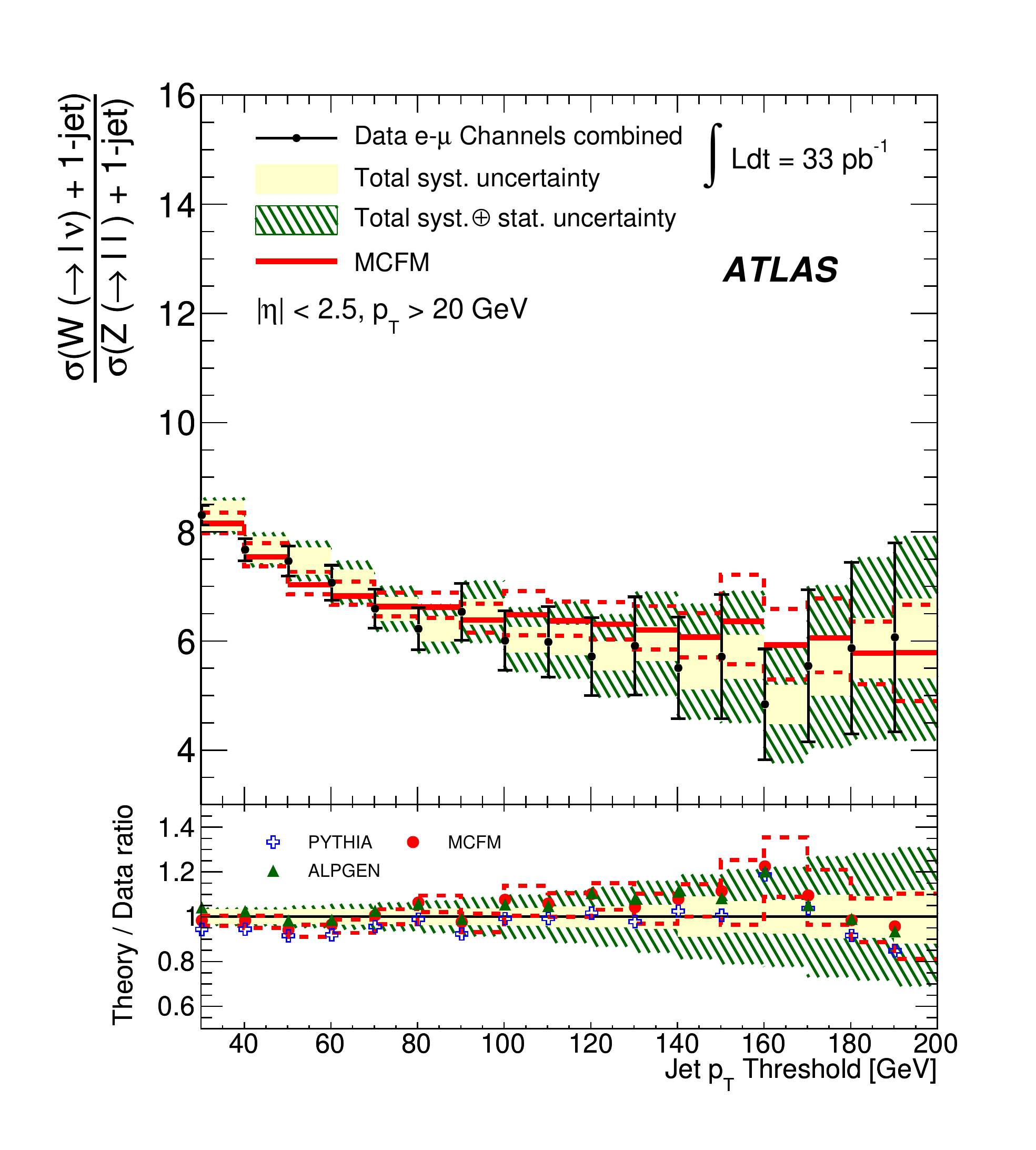}
  \end{center}
  \end{minipage}
  \begin{minipage}{0.49\linewidth}
  \begin{center}
    \includegraphics[width=0.95\linewidth]{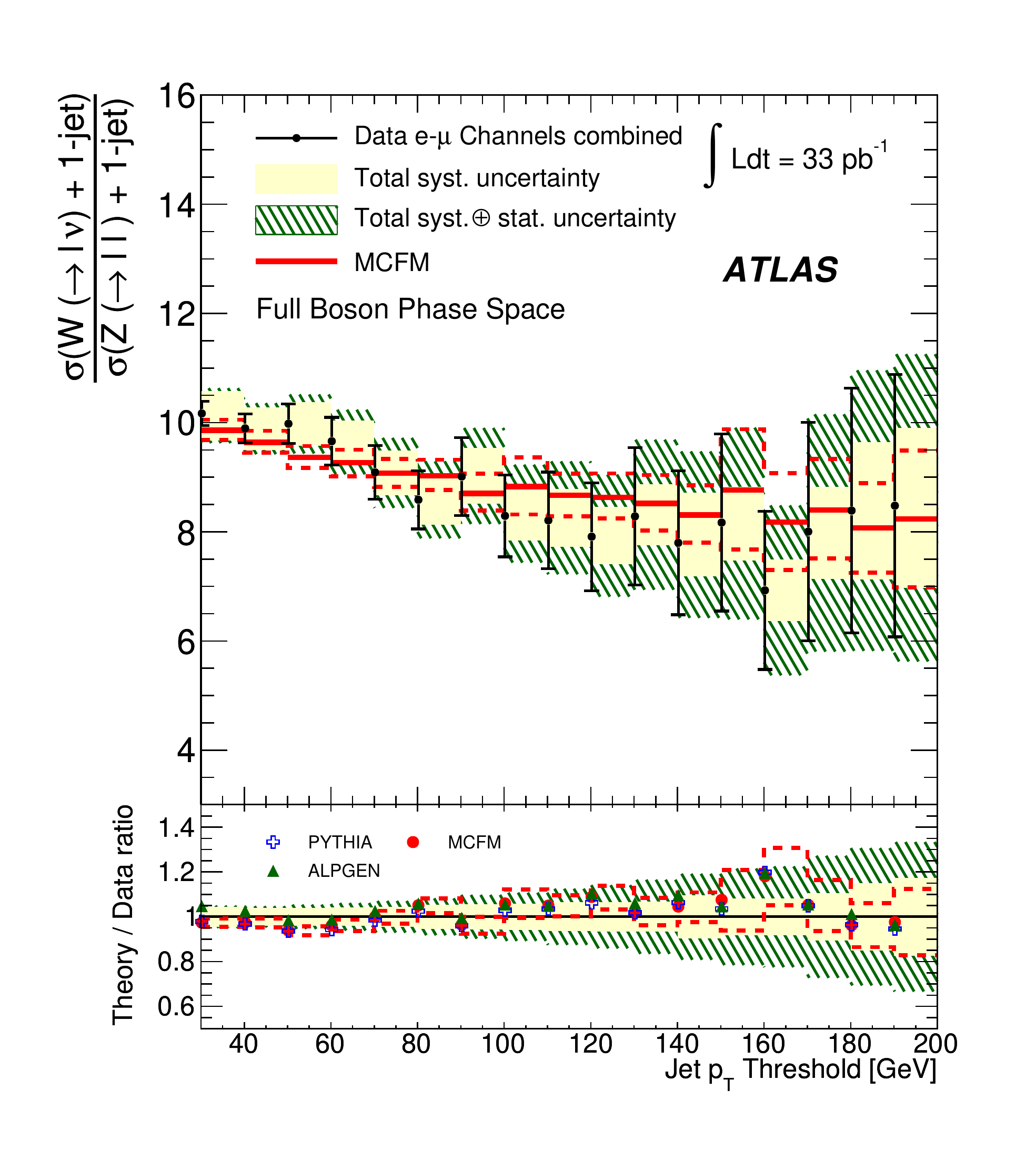}
  \end{center}
  \end{minipage}
\caption{
Left: Combined electron and muon results for \rjets in a common fiducial
region.  The results are compared to predictions from MCFM (corrected to
particle level).  Data are shown with black error bars indicating the
statistical uncertainties. The yellow band shows all systematic
uncertainties added in quadrature and the green band shows statistical and
systematic uncertainties added in quadrature.  The theory uncertainty
(dashed line) includes contributions from PDF and renormalisation and
factorization scales.  Right: Combined electron and muon results for \rjets
extrapolated to the total phase space.  Note that these threshold data and
their associated uncertainties are correlated between bins.}
\label{fig:combination}
\end{figure*}

Electron and muon channel results were found to be compatible and therefore
combined to reduce the statistical and uncorrelated systematic uncertainties
on the result. Each channel was extrapolated to a common phase space,
defined as $|\eta_\ell| < 2.5$ before any QED radiation (Born level) with
\pythia.  The electron channel was further
corrected for the effect of the electron-jet isolation requirements on the
acceptance.  This extrapolation to a common fiducial region decreases the
value of the ratio for both channels primarily due to the more central
distribution of leptons from the Z.  The results were combined using a
Bayesian approach~\cite{Caldwell:2010zz} in the combination of systematic
uncertainties accounting for correlations between them.  The systematic
uncertainties from \met, jet energy scale and resolution and electroweak
background sources were considered fully correlated between the electron and
muon channels.  The combined result is shown in
Figure~\ref{fig:combination}~(left).  The value of \rjets for the
lowest jet \pt threshold of 30 \GeV was found to be $8.29 \pm 0.18\,\mathrm{(stat)} \pm 0.28 \,\mathrm{(syst)}$. This combined
measurement was also extrapolated to the full phase space, as shown in
Figure~\ref{fig:combination}~(right). For a jet \pt threshold of 30 \GeV
this ratio was found to be $10.13 \pm 0.22 \,\mathrm{(stat)} \pm 0.45 \,\mathrm{(syst)}$.

\section{Summary}
We present a measurement of the ratio of the production
cross sections of the gauge bosons \wboson and \zboson in association with
exactly one jet, as a function of jet \pt threshold. This ratio was
measured in fiducial phase space of the detector separately for muons
and electrons. These results were also extrapolated to a common phase
space and combined, as well as extrapolated to be the full phase space
of the boson decay products. These results were
provided as a function of jet \pt threshold from 30 to 200 \GeV, exploring
the transition region of electroweak scale breaking in perturbative jet
production. The accepted theoretical model was found to be consistent
with all results. This measurement has the advantage of having reduced
theoretical and experimental systematic uncertainties due to correlations
between the \wboson and \zboson processes. This measurement builds the foundations
of a high precision test of the Standard Model, and provides
model-independent sensitivity to new physics coupling to leptons and
jets.  Comparisons with LO and NLO perturbative QCD predictions were
made and found to be in agreement with data over the jet \pt threshold
range covered by this measurement.

\bigskip 



\bibliography{rjets_dpf2011}

\end{document}